\documentclass[
  reprint,
  aps,
  prapplied,
  amsmath,
  amssymb,
  floatfix,
  longbibliography,
  nobibnotes
]{revtex4-2}

\usepackage{graphicx}
\usepackage{dcolumn}
\usepackage{bm}
\usepackage{hyperref}
\usepackage{xcolor}

\begin{document}

% =========================
% Title information
% =========================
\title{Geometric estimation of NV charge-state contributions from a low-dimensional spectral representation}

\author{Yuto Yamakawa}
\author{Keisuke Oshimi}
\author{Keigo Arai}
\email{arai.k.835f@m.isct.ac.jp}

\affiliation{
School of Engineering, Institute of Science Tokyo,
Yokohama, Kanagawa 226-8501, Japan
}

%\date{\today}

% =========================
% Abstract
% =========================
\begin{abstract}
Nitrogen-vacancy (NV) centers in diamond exist in neutral (NV$^0$) and negatively charged (NV$^-$) states, and quantifying their respective photoluminescence (PL) contributions is important for charge-state-based measurements. Existing methods either require additional experimental control or may suffer from limited physical identifiability. Here, we introduce a geometric method for determining NV charge-state contributions from PL spectra acquired at different excitation intensities. Each spectrum is mapped into a low-dimensional space through broadband spectral weighting, where changes in the relative NV$^0$ and NV$^-$ contributions trace a one-dimensional trajectory. Zero-phonon-line information from two spectra physically calibrates this trajectory, enabling the NV$^-$ PL contribution to be determined by geometric projection. Measurements on two bulk single-crystal diamond samples yielded NV$^-$ contributions in close agreement with those obtained using an independent dual-excitation reference-spectrum method, with root-mean-square errors of 1.11 and 0.31 percentage points. Under additive spectral noise, the proposed method exhibited more than an order of magnitude less variation than a ZPL-only method. Fisher-information analysis further showed that the three-dimensional CIE XYZ representation retained approximately 81\% of the information about the NV$^{-}$ contribution available in the full spectrum. These results establish a physically calibrated approach to low-dimensional spectral estimation that combines reduced experimental overhead with robust parameter estimation and provides a general framework for spectral sensing governed by a small number of physical degrees of freedom.
\end{abstract}

\maketitle

% =========================
% 1. Introduction
% =========================
\section{Introduction}
The nitrogen-vacancy (NV) center in diamond is a widely used solid-state quantum defect for quantum sensing under ambient conditions \cite{Doherty2013,Degen2017}. Its spin-dependent optical response enables measurements of magnetic fields \cite{Balasubramanian2008,Grinolds2013,Rondin2014}, temperature \cite{Kucsko2013}, electric fields \cite{Dolde2011}, and other physical quantities. NV centers exist primarily in the neutral (NV$^0$) and negatively charged (NV$^-$) states \cite{Aslam2013}, with NV$^{-}$ providing the spin-dependent optical transitions commonly used for quantum sensing. Charge-state conversion between NV$^0$ and NV$^-$ is also sensitive to the local electrical and chemical environment, including surface conditions, pH, and charged molecules \cite{Rondin2010,Yuan2020,Sow2020,Kremarov2021}, and can be actively controlled by optical excitation \cite{Gao2022}. Quantitative determination of the NV charge state is therefore important both for characterizing NV-based sensors and for emerging schemes that use the charge-state response itself to probe local electrostatic and electrochemical environments \cite{Kimura2026}.

Photoluminescence (PL) spectroscopy provides direct optical signatures of NV$^0$ and NV$^-$ through their distinct zero-phonon lines (ZPLs) and emission spectral profiles \cite{Alsid2019}. However, the measured PL spectrum is a superposition of emission from the two charge states, whose broad phonon sidebands substantially overlap. Quantitatively separating their contributions from a measured spectrum therefore remains challenging.

Several approaches have been developed to estimate NV charge-state contributions from PL spectra. Reference-spectrum methods construct sample-specific NV$^0$ and NV$^-$ spectra by introducing additional experimental controls, such as excitation wavelength, magnetic field, or microwave irradiation \cite{Thalassinos2025, Chakraborty2022, AudeCraik2020}. These methods provide physically interpretable estimates but require additional measurements and control modalities. ZPL-based approaches offer a simpler alternative by converting the measured NV$^0$ and NV$^-$ ZPL intensities into total PL contributions using Debye--Waller factors \cite{Zhao2012,Acosta2009}, but their quantitative accuracy depends on literature-derived emission parameters and on local spectral information that can be sensitive to noise. Data-driven decomposition methods such as non-negative matrix factorization (NNMF) can instead operate on a series of spectra measured while varying a single control parameter \cite{Savinov2022}. However, accurate spectral reconstruction does not by itself ensure unique physical identification of the underlying components \cite{Raabova2026, Laurberg2008, AkbariLakeh2022, GillisRajko2023}. A remaining challenge is therefore to obtain sample- and measurement-system-specific charge-state estimates without requiring additional experimental control while retaining explicit physical calibration.

Here, we propose a geometric method for estimating the NV$^-$ PL contribution from a series of spectra obtained by varying only the excitation intensity. If the spectral shapes of NV$^0$ and NV$^-$ remain fixed, their mixtures are governed by a single mixing fraction and therefore form a one-dimensional trajectory under any linear spectral mapping. The original spectral-decomposition problem can thus be reformulated as a geometric problem of determining the position of each spectrum along this trajectory. As a standardized and training-free implementation, we use the Commission Internationale de l'\'{E}clairage (CIE) 1931 XYZ color-matching functions \cite{CIE2018} to map the broadband spectra into a three-dimensional space. ZPL information from two spectra then provides a physical calibration between position along the trajectory and the NV$^-$ PL contribution, without requiring literature-derived Debye--Waller factors or separately measured reference spectra. We validate the method on two bulk single-crystal diamond samples against an independent dual-excitation reference-spectrum method and evaluate its robustness to spectral noise. We further quantify the parameter information retained by the low-dimensional representation using Fisher information and compare the geometric framework with data-driven spectral decomposition. This analysis distinguishes spectral dimensionality, reconstruction accuracy, and physically identifiable parameter information as separate properties of spectral representations.

% =========================
% 2. Method
% =========================
\section{Methods}

\subsection{Experimental setup and PL measurements}

Two bulk single-crystal diamond samples were investigated: Sample A, a commercially available diamond (DNVB14, Thorlabs, Inc.), and Sample B, a diamond fabricated by Diancy Co. The samples were excited using a continuous-wave 532~nm laser (MLL-S-532B-300mW, CNI Laser). The nominal laser-output setting was varied from 1 to 100~mW (1, 2, 5, 7, 10, 20, 50, 70, and 100~mW) using the laser controller. These values refer to the laser-output settings rather than the optical power incident on the sample. A reflective neutral-density filter (ND10A, Thorlabs; OD = 1.0) was placed immediately after the laser. The excitation beam was focused onto the sample through a 100$\times$ objective lens (MPLFLN100X, Evident/Olympus; NA = 0.90), which was also used to collect the resulting photoluminescence. Before entering the spectrometer, the collected light was attenuated using a second identical reflective neutral-density filter. The PL spectra were recorded using a spectrometer (SpectraPro SP-2150, Teledyne Princeton Instruments) equipped with a CCD camera (PIXIS: 256E, Teledyne Princeton Instruments).

\subsection{Geometric quantification of charge-state-dependent photoluminescence}

\begin{figure*}[htbp]
  \centering
  \includegraphics[width=4.50in]{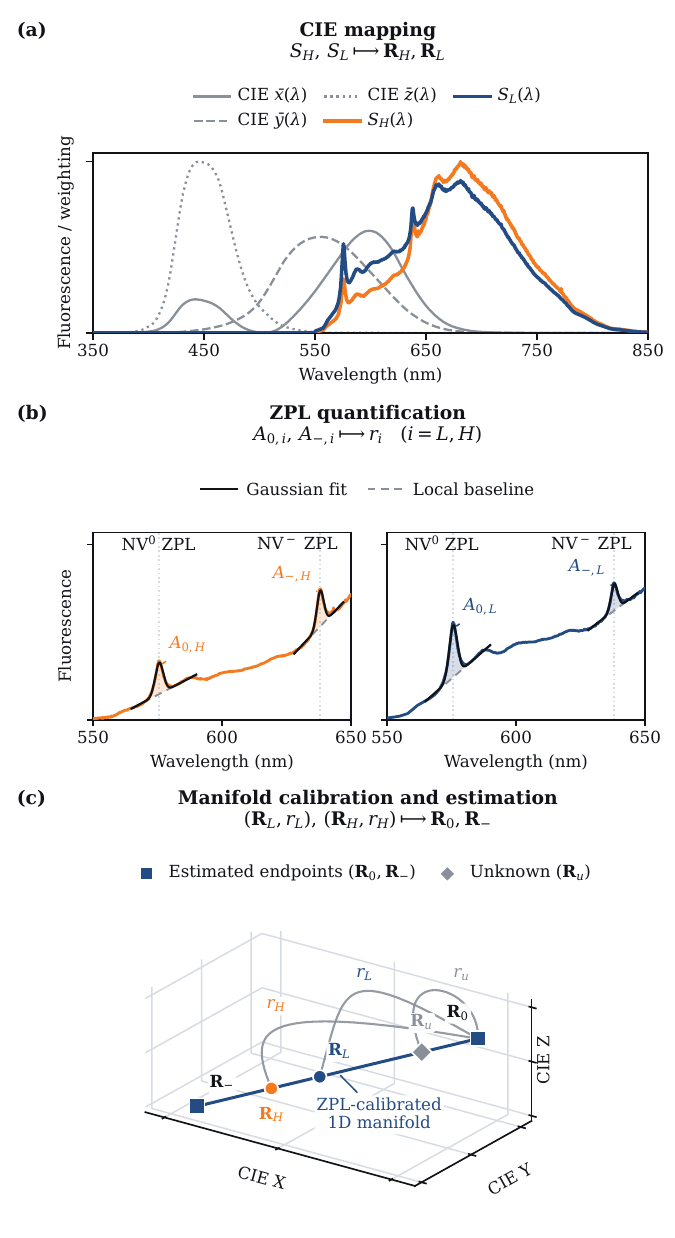}
  \caption{
  Geometric quantification of charge-state-dependent PL.
  (a) Two area-normalized PL spectra, $S_{\mathrm{H}}(\lambda)$ and $S_{\mathrm{L}}(\lambda)$, with higher and lower NV$^-$ contributions, respectively.
  Each spectrum is mapped to a vector, $\mathbf{R}_{\mathrm{H}}$ or $\mathbf{R}_{\mathrm{L}}$, in CIE XYZ space by integration over 550--850~nm using the CIE 1931 color-matching functions $\bar{x}(\lambda)$, $\bar{y}(\lambda)$, and $\bar{z}(\lambda)$.
  (b) Gaussian fitting of the NV$^0$ and NV$^-$ ZPLs yields the four ZPL areas $A_{0,\mathrm{H}}$, $A_{-,\mathrm{H}}$, $A_{0,\mathrm{L}}$, and $A_{-,\mathrm{L}}$, from which the NV$^-$ PL contribution ratios $r_{\mathrm{H}}$ and $r_{\mathrm{L}}$ are determined.
  (c) The vectors $\mathbf{R}_{\mathrm{H}}$ and $\mathbf{R}_{\mathrm{L}}$ from (a), together with $r_{\mathrm{H}}$ and $r_{\mathrm{L}}$ from (b), determine the pure-state endpoints $\mathbf{R}_0$ and $\mathbf{R}_-$. The line connecting these endpoints defines the one-dimensional spectral manifold, onto which an unknown vector $\mathbf{R}_{\mathrm{u}}$ is projected to geometrically estimate its NV$^-$ PL contribution ratio $r_{\mathrm{u}}$.
  }
  \label{fig:cie_zpl_workflow}
\end{figure*}

The NV$^-$ PL contribution ratio is defined as $r\equiv P_-/(P_0+P_-)$, where $P_0$ and $P_-$ denote the total PL contributions from NV$^0$ and NV$^-$, respectively. The workflow of the proposed method is illustrated in Fig.~\ref{fig:cie_zpl_workflow}.

When the emission spectral shapes of NV$^0$ and NV$^-$ remain unchanged and the measured spectral variation arises from changes in their relative PL contributions, the spectral variation can be described by a single degree of freedom, the NV$^-$ PL contribution ratio $r$. Each PL spectrum is normalized to unit area over the analyzed wavelength range $\Lambda$, and $S_0(\lambda)$ and $S_-(\lambda)$ denote the area-normalized basis spectra of NV$^0$ and NV$^-$, respectively. Under the two-component mixing model, the area-normalized spectrum for a given NV$^-$ PL contribution ratio $r$ is expressed as
\begin{equation}
S(\lambda;r)
=
(1-r)S_0(\lambda)+rS_-(\lambda).
\label{eq:spectral_mixture}
\end{equation}

To represent this one-degree-of-freedom spectral variation geometrically, each spectrum is linearly mapped into a low-dimensional space using multiple weighting functions. Defining the vector-valued weighting function as $\mathbf{w}(\lambda)=[w_1(\lambda),w_2(\lambda),\ldots,w_K(\lambda)]^{\mathrm{T}}$, the low-dimensional representation of a spectrum $S(\lambda)$ is given by
\begin{equation}
\mathbf{R}[S]
\equiv
\int_{\Lambda}
S(\lambda)\mathbf{w}(\lambda)\,d\lambda.
\label{eq:weighted_spectral_mapping}
\end{equation}
The vectors corresponding to the NV$^0$ and NV$^-$ basis spectra are denoted by $\mathbf{R}_0\equiv\mathbf{R}[S_0]$ and $\mathbf{R}_-\equiv\mathbf{R}[S_-]$, respectively. Applying this linear mapping to the two-component mixing model in Eq.~\eqref{eq:spectral_mixture} gives
\begin{equation}
\mathbf{R}(r)
=
(1-r)\mathbf{R}_0+r\mathbf{R}_-.
\label{eq:low_dimensional_linear}
\end{equation}
Accordingly, variation in $r$ is represented as a change in position along the line connecting $\mathbf{R}_0$ and $\mathbf{R}_-$ in the low-dimensional space. This one-dimensional affine trajectory is hereafter referred to as the spectral manifold. The original high-dimensional spectral mixing problem is thereby transformed into a geometric problem of determining the position along this line. In the present implementation, $K=3$, and the CIE 1931 color-matching functions are used as the weighting functions \cite{CIE2019CMF}, $\mathbf{w}(\lambda)=[\overline{x}(\lambda),\overline{y}(\lambda),\overline{z}(\lambda)]^{\mathrm{T}}$, so that the components of $\mathbf{R}=(X,Y,Z)^{\mathrm{T}}$ correspond to the CIE XYZ coordinates. This choice provides a fixed and standardized broadband projection and avoids introducing a dataset-dependent basis into the geometric calibration.

However, determining the NV$^-$ PL contribution ratio $r$ from a position along the spectral manifold requires knowledge of the pure NV$^0$ and NV$^-$ endpoints, $\mathbf{R}_0$ and $\mathbf{R}_-$, corresponding to $r=0$ and $r=1$, respectively. The endpoints are determined using the ZPL information contained in two measured spectra, $S_{\mathrm{L}}(\lambda)$ and $S_{\mathrm{H}}(\lambda)$, because the pure-state spectra are not directly observed in the present measurement series. Here, $S_{\mathrm{L}}(\lambda)$ and $S_{\mathrm{H}}(\lambda)$ denote the spectra with lower and higher NV$^-$ contributions, respectively. Their NV$^-$ PL contribution ratios are denoted by $r_{\mathrm{L}}$ and $r_{\mathrm{H}}$, and the corresponding NV$^0$ and NV$^-$ ZPL areas are denoted by $A_{0,\mathrm{L}}$, $A_{-,\mathrm{L}}$, $A_{0,\mathrm{H}}$, and $A_{-,\mathrm{H}}$. A sample- and measurement-system-specific coefficient $\alpha$ accounts for the different ZPL fractions of the total NV$^0$ and NV$^-$ PL. Because the two calibration spectra are area normalized, the total NV$^0$ and NV$^-$ contributions are constrained to sum to unity. Requiring the same conversion coefficient $\alpha$ to describe both spectra therefore allows $\alpha$ to be determined directly from their measured ZPL areas. The coefficient $\alpha$ and the PL contribution ratios of the two calibration spectra are determined from the measured ZPL areas as
\begin{equation}
\begin{aligned}
\alpha
&=
\frac{
A_{-,\mathrm{H}}-A_{-,\mathrm{L}}
}{
A_{0,\mathrm{L}}-A_{0,\mathrm{H}}
},
\\
r_i
&=
\frac{
A_{-,i}
}{
\alpha A_{0,i}+A_{-,i}
},
\qquad
i\in\{\mathrm{L},\mathrm{H}\}.
\end{aligned}
\label{eq:zpl_calibration}
\end{equation}
Together with the corresponding mapped vectors $\mathbf{R}_{\mathrm{L}}$ and $\mathbf{R}_{\mathrm{H}}$, the ZPL-derived ratios $r_{\mathrm{L}}$ and $r_{\mathrm{H}}$ define two calibration points, $(\mathbf{R}_{\mathrm{L}},r_{\mathrm{L}})$ and $(\mathbf{R}_{\mathrm{H}},r_{\mathrm{H}})$. Substituting these two calibration points into Eq.~\eqref{eq:low_dimensional_linear} determines the endpoint vectors $\mathbf{R}_0$ and $\mathbf{R}_-$.

Once the spectral manifold and the endpoints corresponding to $r=0$ and $r=1$ have been determined, the NV$^-$ PL contribution ratio of an unknown spectrum can be obtained by projecting its low-dimensional vector $\mathbf{R}_{\mathrm{u}}$ onto the spectral manifold defined by $\mathbf{R}_0$ and $\mathbf{R}_-$. Consistent with the linear relation in Eq.~\eqref{eq:low_dimensional_linear}, the estimated NV$^-$ PL contribution ratio is
\begin{equation}
r_{\mathrm{u}}
=
\frac{
(\mathbf{R}_{\mathrm{u}}-\mathbf{R}_0)
\cdot
(\mathbf{R}_--\mathbf{R}_0)
}{
\left\|
\mathbf{R}_--\mathbf{R}_0
\right\|^2
}.
\label{eq:geometric_projection}
\end{equation}
Thus, the NV$^-$ PL contribution ratio of an unknown spectrum is obtained directly from its projected position along the calibrated spectral manifold. This implementation, which combines CIE XYZ mapping with ZPL-based calibration, is hereafter referred to as the CIE--ZPL method.

\subsection{Comparison methods for charge-state estimation}

To evaluate the quantitative validity of the proposed method, the NV$^-$ PL contribution ratios were also determined using the dual-excitation protocol (DEP) \cite{Thalassinos2025}, which provides an independent, sample- and measurement-system-specific spectral-decomposition method. For the DEP measurements, 405~nm excitation was provided by a continuous-wave laser module (PL255, Thorlabs). DEP constructs NV$^0$ and NV$^-$ reference spectra from PL measurements performed under different excitation wavelengths. In the present analysis, the spectrum measured under 405~nm excitation was used as the NV$^0$ reference spectrum. The 405 and 532~nm spectra were normalized to the same NV$^0$ contribution over the 550--600~nm range, and their difference was used to derive the NV$^-$ reference spectrum. Each spectrum in the 532~nm excitation-power series was subsequently fitted as a linear combination of the two reference spectra to determine the NV$^-$ PL contribution ratio. Further details of the reference-spectrum construction and fitting procedure are provided in the Supplemental Material below.

For comparison with a conventional ZPL-based approach, the NV$^-$ PL contribution ratios were also estimated from the fitted NV$^0$ and NV$^-$ ZPL areas using literature-derived Debye--Waller factors, hereafter referred to as the DWF--ZPL method. In this method, the fitted ZPL areas were converted into the corresponding total PL contributions by accounting for the fractions of the NV$^0$ and NV$^-$ emissions contained in their respective ZPLs.

The agreement between the CIE--ZPL and DEP estimates was quantified using the root-mean-square error,
\begin{equation}
\mathrm{RMSE}
=
\left[
\frac{1}{M}
\sum_{i=1}^{M}
\left(
r_{\mathrm{CIE},i}
-
r_{\mathrm{DEP},i}
\right)^2
\right]^{1/2},
\end{equation}
where $M$ is the number of nominal laser-output settings, and
$r_{\mathrm{CIE},i}$ and $r_{\mathrm{DEP},i}$ denote the NV$^-$ PL contribution ratios estimated by CIE--ZPL and DEP, respectively, at the $i$th nominal laser-output setting.

\subsection{Evaluation of estimation precision and Fisher information}

To evaluate the robustness of charge-state estimation to spectral noise after calibration, we performed a noise-perturbation analysis. Only the spectrum to be estimated was varied, while all method-specific calibration inputs were held fixed. Specifically, the endpoint vectors $\mathbf{R}_0$ and $\mathbf{R}_-$ for CIE--ZPL, the NV$^0$ and NV$^-$ reference spectra for DEP, and the literature-derived parameters for DWF--ZPL were fixed throughout the analysis. A representative measured PL spectrum was then perturbed with additive white Gaussian noise, and the NV$^-$ PL contribution ratio was repeatedly estimated using each method. The noise-induced variation in the estimated ratio was quantified by its standard deviation. Further details of the noise model and simulation conditions are provided in the Supplemental Material below.

To complement the Monte Carlo evaluation, Fisher information (FI) was used to quantify the information about $r$ retained at different stages of spectral processing. FI was evaluated for three observation models: the full spectrum, the CIE XYZ representation, and the ZPL observables used for ZPL-based estimation. For the CIE representation, the covariance induced by area normalization of the noisy spectrum was explicitly propagated through the normalization step and subsequent CIE mapping. The resulting covariance matrix was used to calculate the FI and corresponding Cramér--Rao lower bound (CRLB) for $r$. The analytical covariance propagation was independently validated against Monte Carlo simulations. Full derivations, noise-model definitions, and validation procedures are provided in the Supplemental Material below.

OpenAI ChatGPT (GPT-5.6 Sol) was used as an assistive tool in portions of the mathematical and computational analysis, including the examination of analytical derivations, development and debugging of Python analysis code, and verification of numerical procedures. Task-specific instructions were provided together with the relevant analytical definitions, source data, and analysis code as context. All AI-assisted derivations, code modifications, numerical results, and scientific interpretations included in this work were reviewed and verified against the analytical definitions, source data, and reproducible analysis code.

% =========================
% 3. Results & Discussion
% =========================

\section{Results and Discussion}

\begin{figure*}[htbp]
  \centering
  \includegraphics[width=6.69in]{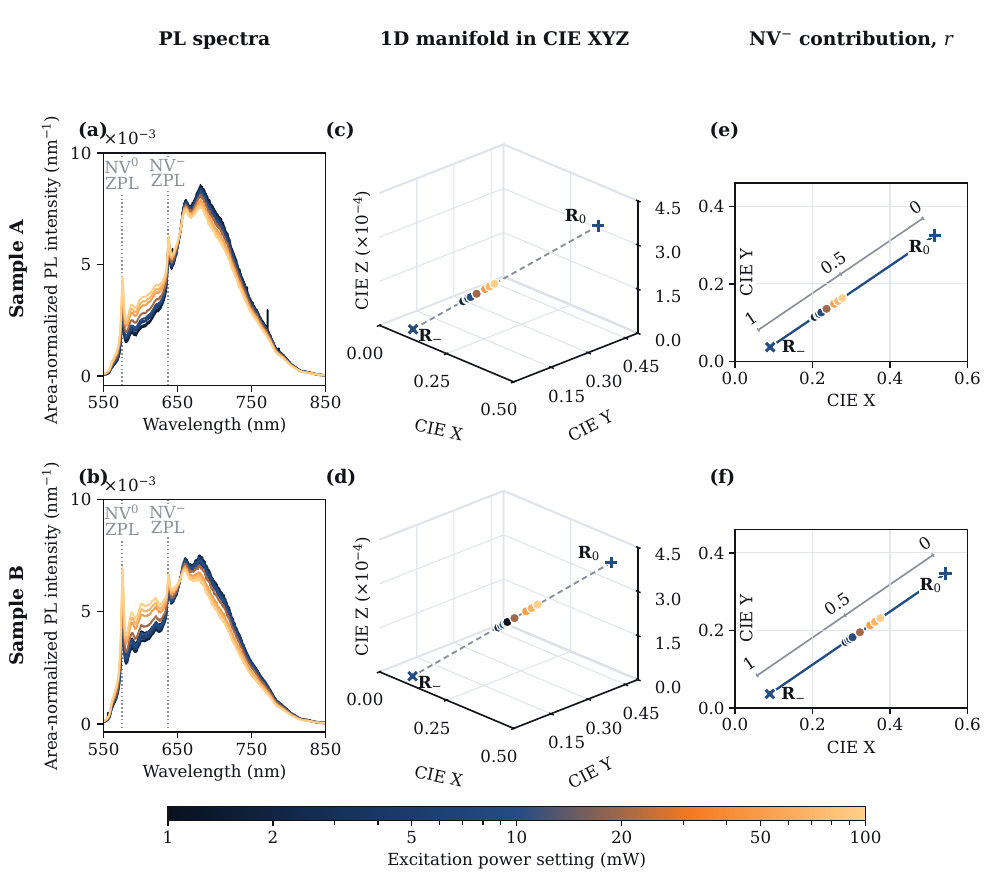}
  \caption{
  Application of the CIE--ZPL method to excitation-power-dependent PL spectra.
  (a,b) Area-normalized PL spectra acquired at different nominal laser-output settings for Samples A and B, respectively.
  The dotted vertical lines indicate the NV$^0$ and NV$^-$ ZPL wavelengths.
  (c,d) Corresponding CIE XYZ vectors. The estimated pure-state endpoints are indicated by the endpoint markers, and the line connecting them represents the calibrated one-dimensional spectral manifold.
  (e,f) Projections of the measured spectra onto the calibrated spectral manifolds, displayed in the CIE XY plane. The displaced parallel axis indicates the estimated NV$^-$ PL contribution ratio $r$, with $r=0$ and $r=1$ corresponding to the NV$^0$ and NV$^-$ endpoints, respectively.
  }
  \label{fig:cie_spectra_xyz}
\end{figure*}

To experimentally evaluate the proposed geometric framework, we applied the CIE--ZPL method to excitation-power-dependent PL spectra obtained from Samples A and B. As shown in Figs.~\ref{fig:cie_spectra_xyz}(a) and \ref{fig:cie_spectra_xyz}(b), the PL spectral shapes of both samples changed systematically with excitation power. When these spectra were mapped into CIE XYZ space, the resulting vectors formed approximately one-dimensional trajectories for both samples, as shown in Figs.~\ref{fig:cie_spectra_xyz}(c) and \ref{fig:cie_spectra_xyz}(d). The NV$^0$ and NV$^-$ endpoint vectors, $\mathbf{R}_0$ and $\mathbf{R}_-$, defined the calibrated one-dimensional spectral manifold, onto which each measured vector was projected according to Eq.~\eqref{eq:geometric_projection}. The projected positions shown in Figs.~\ref{fig:cie_spectra_xyz}(e) and \ref{fig:cie_spectra_xyz}(f) yielded estimates of the NV$^-$ PL contribution ratio for each measured spectrum. Thus, the CIE mapping provided a low-dimensional coordinate representation of the observed spectral variation, while the ZPL-based calibration and geometric projection converted position along the trajectory into the NV$^-$ PL contribution ratio.

Importantly, the observed one-dimensionality is not introduced by the CIE mapping. Principal component analysis of the original PL spectra showed that the first principal component accounted for 99.81\% and 99.83\% of the spectral variance for Samples A and B, respectively. The spectral variation is therefore already dominated by a single degree of freedom in the full spectral space. The CIE mapping provides a fixed coordinate representation of this low-dimensional variation, whereas the subsequent ZPL-based calibration relates position along the manifold to the physical parameter $r$. Thus, the physical dimensionality of the spectral evolution and the choice of coordinates used to represent it are conceptually distinct.

We also examined whether the same spectral series could be uniquely decomposed using NNMF. Standard random initialization and an alternative initialization based on the measured 1 and 100 mW spectra yielded distinct spectral components and NV$^-$ contribution ratios despite producing comparable reconstruction errors (see Supplemental Material below). This result illustrates that accurate reconstruction of a low-dimensional spectral dataset does not necessarily guarantee physical identifiability of the underlying components. In the present approach, physical identifiability is instead introduced explicitly through ZPL-based calibration of the position along the spectral manifold.

\begin{figure}[htbp]

    \centering

    \includegraphics[width=3.37in]{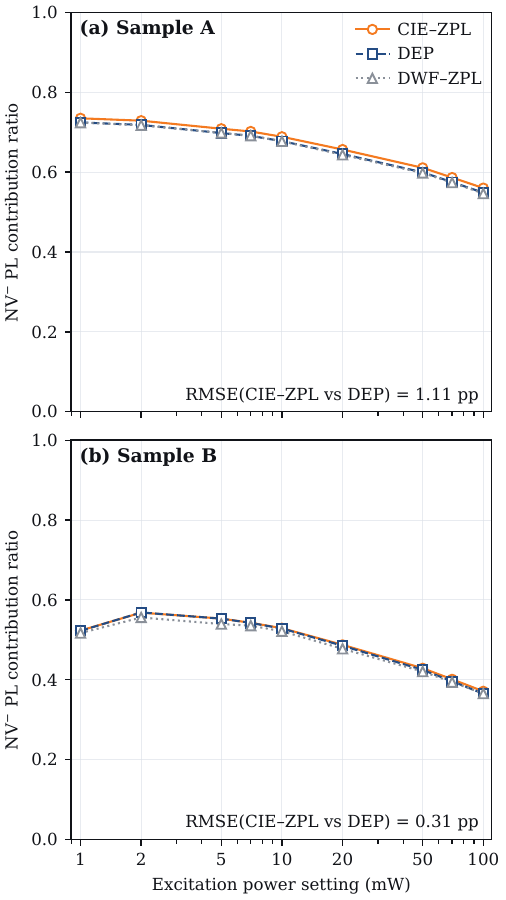}

    \caption{
    Comparison of three NV$^-$ PL contribution-ratio estimation methods.
    NV$^-$ PL contribution ratios were estimated as a function of excitation power for (a) Sample A and (b) Sample B using the CIE--ZPL, DEP, and DWF--ZPL methods.
    The RMSEs between the CIE--ZPL and DEP estimates were 1.11 and 0.31 percentage points (pp) for Samples A and B, respectively.
    }

    \label{fig:cie_vs_established}

\end{figure}

Figure~\ref{fig:cie_vs_established} compares the NV$^-$ PL contribution ratios obtained using the CIE--ZPL, DEP, and DWF--ZPL methods. For both Samples A and B, the CIE--ZPL estimates closely followed those obtained using DEP, with RMSEs of 1.11 and 0.31 percentage points, respectively. The DWF--ZPL method also yielded a similar excitation-power dependence under the present experimental conditions, indicating that the literature-derived Debye--Waller factors yielded compatible estimates for the present samples and measurement configuration. However, this agreement does not establish the transferability of these literature-derived factors to samples with different emission characteristics or optical systems with different wavelength-dependent responses. Thus, the close agreement between the CIE--ZPL and independently obtained DEP estimates supports the quantitative validity of the proposed CIE--ZPL method under the investigated conditions.

\begin{figure*}[htbp]
    \centering
    \includegraphics[width=6.69in]{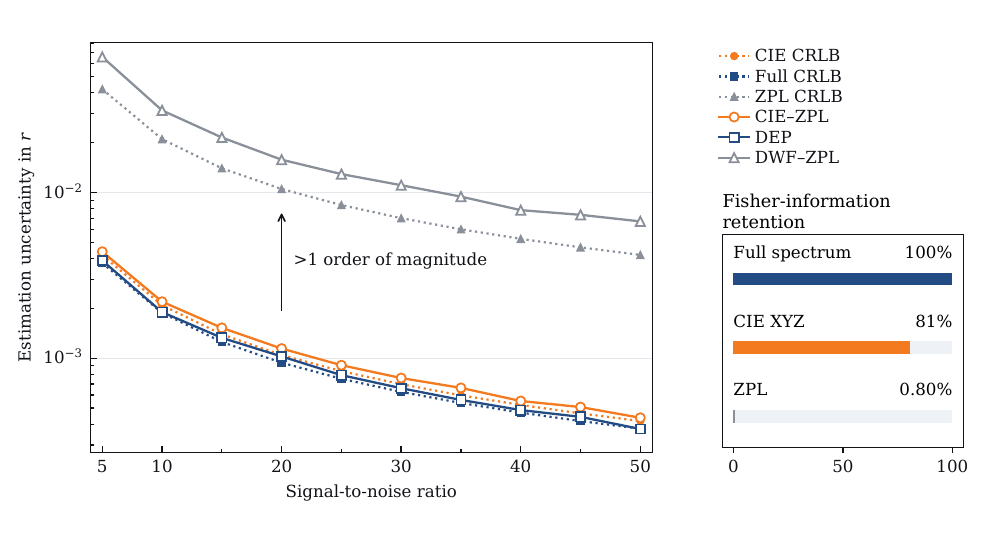}
    \caption{
    Comparison of the precision of three NV$^-$ PL contribution-ratio estimation methods under additive spectral noise and the corresponding Fisher-information-based CRLBs.
    Zero-mean white Gaussian noise was added to the spectrum to be estimated while all method-specific calibration inputs were held fixed, and the SNR was varied to compare the CIE--ZPL, DEP, and DWF--ZPL methods.
    The standard deviation of the estimated NV$^-$ PL contribution ratio is shown together with the CRLBs calculated for the full-spectrum (Full), CIE XYZ (CIE), and ZPL-observable (ZPL) observation models under the same noise assumption.
    CIE--ZPL exhibited estimation precision comparable to DEP, with standard deviations more than one order of magnitude smaller than those of DWF--ZPL. Inset: Fisher information retained by the full spectrum, CIE XYZ representation, and ZPL observables, normalized to that of the full spectrum.
    }
    \label{noise_robustness}
\end{figure*}

Figure~\ref{noise_robustness} shows the noise-induced estimation uncertainty of the three methods and the Fisher-information-based CRLBs for the full-spectrum, CIE, and ZPL observation models. Across the investigated signal-to-noise ratio (SNR) range, CIE--ZPL and DEP exhibited substantially smaller estimation uncertainties than DWF--ZPL. The standard deviation obtained using DWF--ZPL was approximately 14--15 times that obtained using CIE--ZPL and approximately 15--18 times that obtained using DEP. These results demonstrate that CIE--ZPL provides post-calibration estimation precision close to that of DEP and substantially higher than that of DWF--ZPL.

The CRLBs showed the same overall ordering and SNR dependence as the Monte Carlo results, indicating that spectral processing changes the amount of information retained about $r$. Relative to the full spectrum, the CIE and ZPL representations retained approximately 81\% and 0.80\% of the Fisher information, respectively. Notably, this level of information retention is achieved using only three standardized weighting functions that were not designed or optimized for NV charge-state estimation. The remaining information loss in the CIE representation is consistent with the spectral mismatch between the CIE weighting functions and the measured PL spectrum. In particular, the CIE color-matching functions place relatively weak weight on the long-wavelength portion of the NV emission spectrum. The larger DWF--ZPL uncertainty compared with the ZPL CRLB may arise from the nonlinear conversion of the two fitted ZPL areas to $r$.

The scope of this analysis is limited to the estimation stage with all method-specific calibration inputs held fixed. It does not include uncertainty arising from the ZPL fitting used to determine $\alpha$ and the endpoint vectors $\mathbf{R}_0$ and $\mathbf{R}_-$ in CIE--ZPL, or from the measurements used to construct the DEP reference spectra. Errors in the calibration spectra may consequently propagate through the endpoint vectors to subsequent estimates. A possible extension would be to determine the calibration parameters using multiple spectra simultaneously by enforcing consistency between the ZPL-derived contribution ratios and the positions of the spectra along the one-dimensional manifold. Such a global calibration may reduce sensitivity to individual calibration spectra.

The geometric framework is not restricted to the CIE XYZ representation. In principle, any linear low-dimensional representation may be used provided that it preserves the spectral variation associated with the two-component mixture. Although the first principal component accounted for 99.8\% of the spectral variance in the original spectra, the CIE XYZ representation retained approximately 81\% of the Fisher information about $r$ available in the full spectrum. This contrast shows that preserving spectral variance and preserving information about the target parameter are distinct criteria \cite{Heavens2000, Beattie2021}. More generally, the existence of a low-dimensional spectral manifold does not imply that an arbitrary low-dimensional representation will preserve the information most relevant to estimating its physical coordinate. The CIE representation is therefore not information-optimal for estimating $r$. Its value in the present work instead lies in providing a standardized, fixed, and training-free broadband projection with a deterministic mapping across datasets. The geometric framework itself is not tied to CIE. Application-specific weighting functions could in principle be designed to retain a larger fraction of the parameter-relevant information without changing the underlying calibration concept \cite{Alsing2018}. The CIE XYZ representation was therefore adopted as a simple and reproducible implementation of the general geometric framework.

An additional practical implication of using a fixed low-dimensional linear projection is that the method need not remain a purely software-based spectral analysis. In principle, the three weighted integrals represented here by the CIE XYZ coordinates could be implemented using a small number of spectrally weighted detection channels, reducing the need for full spectral acquisition after the weighting functions have been fixed. Such hardware-level spectral compression could provide a route toward compact charge-state sensing architectures.

The practical advantage of CIE--ZPL lies not only in its estimation precision but also in the reduced experimental overhead required for calibration. DEP exhibited high estimation precision but requires additional measurements under 405~nm excitation to construct the reference spectra. In contrast, CIE--ZPL is calibrated using only the 532~nm excitation-power series and, after calibration, estimates $r$ by geometric projection of each spectrum. Thus, the method trades additional experimental control for a fixed low-dimensional representation and internal calibration.

The present method assumes a two-component mixing model in which the spectral shapes of NV$^0$ and NV$^-$ remain fixed and only their relative contributions vary. Changes in the spectral shape of either charge state due, for example, to temperature or strain, or the introduction of additional emission components or background can cause the measured spectra to deviate from the one-dimensional manifold. Such deviations therefore define the regime in which the two-component model ceases to be sufficient. Importantly, the off-manifold residual is not only an estimation error but may also serve as a diagnostic for additional physical degrees of freedom, such as temperature- or strain-induced spectral changes, background emission, or contributions from additional optical states.

% =========================
% 5. Conclusion
% =========================
\section{Conclusion}

In this study, we developed a geometric method for quantifying the photoluminescence contribution of negatively charged nitrogen-vacancy centers from excitation-dependent spectra. When the spectra are mapped into a low-dimensional CIE XYZ representation, variations in the relative NV$^0$ and NV$^-$ contributions form an approximately one-dimensional spectral manifold that can be physically calibrated using zero-phonon-line information from only two spectra within the same measurement series. The resulting CIE--ZPL method requires no additional excitation wavelength, literature-derived Debye--Waller factors, or separately measured reference spectra.

For two bulk single-crystal diamond samples, the estimated NV$^-$ contributions closely agreed with those obtained using an independent dual-excitation reference-spectrum method. Under additive spectral noise, CIE--ZPL achieved estimation precision close to the reference-spectrum method and substantially higher than a ZPL-only approach, consistent with the higher FI retention of the broadband low-dimensional representation.

More broadly, our results show that low spectral dimensionality, variance preservation, and information retention for physical-parameter estimation are distinct properties. A low-dimensional representation therefore becomes useful for physical-parameter estimation not simply because it reconstructs the spectra accurately, but because the position along its spectral manifold can be independently calibrated to the target physical quantity. This framework provides a route toward simple, physically interpretable, and potentially hardware-efficient spectral sensing in systems whose spectral evolution is governed by a small number of physical degrees of freedom.

% =========================
% Acknowledgments
% =========================
\section*{Acknowledgments}

This work was supported by JST K Program Japan Grant Number JPMJKP24F3 and JSPS KAKENHI Grant Number JP26K22715.

OpenAI ChatGPT (GPT-5.6 Sol) was also used to assist with literature synthesis and with drafting and revising portions of the scientific text during manuscript preparation. Task-specific instructions were provided together with the relevant literature and scientific context. All cited sources and AI-assisted scientific statements were checked against the original literature and revised as necessary.

% =========================
% Author contributions
% =========================
\section*{Author Contributions}
Yuto Yamakawa: Conceptualization (supporting); Methodology (lead); Investigation (lead); Formal analysis (lead); Software (lead); Visualization (lead); Writing -- original draft (lead); Writing -- review \& editing. Keisuke Oshimi: Methodology (supporting); Writing -- review \& editing. Keigo Arai: Conceptualization (lead); Methodology (supporting); Supervision (lead); Project administration (lead); Funding acquisition (lead); Writing -- review \& editing.

% =========================
% Data Availability
% =========================
\section*{Data Availability}

The data and analysis code that support the findings of this article are openly available in Zenodo~\cite{Yamakawa2026Zenodo}.

% =========================
% References
% =========================
\bibliography{reference}

%%%%%%%%%%%%%%%%%%%%%%% SUPPLEMENTAL MATERIAL %%%%%%%%%%%%%%%%%%%%%%%
\clearpage

% Reset numbering for the Supplemental Material
\setcounter{page}{1}
\setcounter{section}{0}
\setcounter{equation}{0}
\setcounter{figure}{0}
\setcounter{table}{0}

\renewcommand{\thesection}{S\arabic{section}}
\renewcommand{\theequation}{S\arabic{equation}}
\renewcommand{\thefigure}{S\arabic{figure}}
\renewcommand{\thetable}{S\arabic{table}}

\begin{center}
{\LARGE\bfseries Supplemental Material}
\end{center}

\vspace{1em}

\section{Optical setup for PL measurements}

\begin{figure}[htbp]
    \centering
    \includegraphics[width=3.37in]{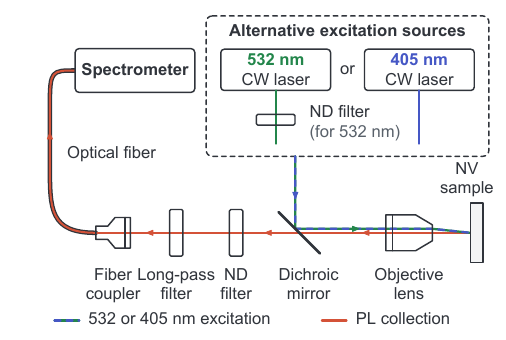}
    \caption{
    Schematic of the PL measurement setup used in this study.
    For the excitation-power-dependent PL measurements, a continuous-wave 532~nm laser was used, with an OD~=~1.0 neutral-density (ND) filter placed immediately after the laser.
    For the DEP measurements, the 532~nm laser was replaced with a continuous-wave 405~nm laser, and the excitation-side ND filter was not used, while the remainder of the optical setup was kept unchanged.
    The excitation light was directed by a dichroic mirror to a 100$\times$ objective lens (NA~=~0.90) and focused onto the NV diamond sample.
    The PL emitted from the sample was collected by the same objective lens and separated from the excitation path by the dichroic mirror.
    The collected PL then passed through an OD~=~1.0 ND filter and a long-pass filter, was coupled into an optical fiber using a fiber coupler, and was finally directed to the spectrometer.
    }
    \label{fig:optical_setup}
\end{figure}

A schematic of the PL measurement setup used in this study is shown in Fig.~\ref{fig:optical_setup}. The same optical setup was used for the 532 and 405~nm measurements, except that the excitation laser was changed and the excitation-side ND filter was used only for 532~nm excitation.

\section{Derivation of the geometric spectral framework}
\label{sec:geometric_framework_derivation}

Let $S_0(\lambda)$ and $S_-(\lambda)$ denote the area-normalized basis spectra of NV$^0$ and NV$^-$, respectively, and let $r\equiv P_-/(P_0+P_-)$ denote the NV$^-$ PL contribution ratio. Under the two-component mixing model, an area-normalized spectrum is expressed as
\begin{equation}
S(\lambda;r)
=
(1-r)S_0(\lambda)+rS_-(\lambda).
\label{eq:si_spectral_mixture}
\end{equation}
Here, $r=0$ and $r=1$ correspond to the pure NV$^0$ and NV$^-$ basis spectra, respectively. This model assumes that the emission spectral shapes of NV$^0$ and NV$^-$ remain unchanged and that the measured spectral variation arises from changes in their relative PL contributions.

To represent this spectral variation in a low-dimensional space, a linear spectral mapping is introduced using the vector-valued weighting function $\mathbf{w}(\lambda)=[w_1(\lambda),w_2(\lambda),\ldots,w_K(\lambda)]^{\mathrm{T}}$. The mapped representation of a spectrum $S(\lambda)$ is defined as
\begin{equation}
\mathbf{R}[S]
\equiv
\int_{\Lambda}
S(\lambda)\mathbf{w}(\lambda)\,d\lambda.
\label{eq:si_linear_mapping}
\end{equation}
Defining $\mathbf{R}_0\equiv\mathbf{R}[S_0]$ and $\mathbf{R}_-\equiv\mathbf{R}[S_-]$, application of the linear mapping to Eq.~\eqref{eq:si_spectral_mixture} gives
\begin{equation}
\mathbf{R}(r)
=
(1-r)\mathbf{R}_0+r\mathbf{R}_-
=
\mathbf{R}_0+r(\mathbf{R}_--\mathbf{R}_0).
\label{eq:si_one_dimensional_manifold}
\end{equation}
Thus, irrespective of the dimension $K$ of the mapped space, the two-component spectral variation is confined to the one-dimensional spectral manifold connecting $\mathbf{R}_0$ and $\mathbf{R}_-$, with the position along the manifold parameterized by the NV$^-$ PL contribution ratio $r$. In the CIE--ZPL implementation used in this work, $K=3$ and $\mathbf{w}(\lambda)=[\overline{x}(\lambda),\overline{y}(\lambda),\overline{z}(\lambda)]^{\mathrm{T}}$, where $\overline{x}(\lambda)$, $\overline{y}(\lambda)$, and $\overline{z}(\lambda)$ are the CIE 1931 color-matching functions, such that $\mathbf{R}=(X,Y,Z)^{\mathrm{T}}$.

To relate the measured ZPL areas to the total PL contribution ratio $r$, let $D_0$ and $D_-$ denote the fractions of the area-normalized NV$^0$ and NV$^-$ basis spectra contained in their respective ZPLs. For a mixed spectrum with contribution ratio $r$, the corresponding ZPL areas are $A_0(r)=D_0(1-r)$ and $A_-(r)=D_-r$. Introducing the coefficient $\alpha\equiv D_-/D_0$ gives
\begin{equation}
r
=
\frac{A_-(r)}
{\alpha A_0(r)+A_-(r)}.
\label{eq:si_zpl_to_r}
\end{equation}
The coefficient $\alpha$ therefore accounts for the different fractions of the total NV$^0$ and NV$^-$ PL contained in their respective ZPLs.

The coefficient $\alpha$ can be determined directly from two area-normalized spectra in the same measurement series without separately determining $D_0$ and $D_-$. Let $S_{\mathrm{L}}(\lambda)$ and $S_{\mathrm{H}}(\lambda)$ denote spectra with lower and higher NV$^-$ contributions, $r_{\mathrm{L}}$ and $r_{\mathrm{H}}$, respectively. Their ZPL areas satisfy $A_{0,i}=D_0(1-r_i)$ and $A_{-,i}=D_-r_i$ for $i\in\{\mathrm{L},\mathrm{H}\}$. The differences between the two calibration spectra are therefore $A_{0,\mathrm{L}}-A_{0,\mathrm{H}}=D_0(r_{\mathrm{H}}-r_{\mathrm{L}})$ and $A_{-,\mathrm{H}}-A_{-,\mathrm{L}}=D_-(r_{\mathrm{H}}-r_{\mathrm{L}})$, yielding
\begin{equation}
\alpha
=
\frac{D_-}{D_0}
=
\frac{
A_{-,\mathrm{H}}-A_{-,\mathrm{L}}
}{
A_{0,\mathrm{L}}-A_{0,\mathrm{H}}
}.
\label{eq:si_alpha}
\end{equation}
Once $\alpha$ has been determined, the NV$^-$ PL contribution ratios of the two calibration spectra are obtained as
\begin{equation}
r_i
=
\frac{
A_{-,i}
}{
\alpha A_{0,i}+A_{-,i}
},
\qquad
i\in\{\mathrm{L},\mathrm{H}\}.
\label{eq:si_calibration_ratios}
\end{equation}
Together with the corresponding mapped vectors $\mathbf{R}_{\mathrm{L}}$ and $\mathbf{R}_{\mathrm{H}}$, the ZPL-derived ratios $r_{\mathrm{L}}$ and $r_{\mathrm{H}}$ define two calibration points, $(\mathbf{R}_{\mathrm{L}},r_{\mathrm{L}})$ and $(\mathbf{R}_{\mathrm{H}},r_{\mathrm{H}})$, on the one-dimensional spectral manifold.

The pure-state endpoint vectors are determined from the two calibration points using $\mathbf{R}_{\mathrm{L}}=(1-r_{\mathrm{L}})\mathbf{R}_0+r_{\mathrm{L}}\mathbf{R}_-$ and $\mathbf{R}_{\mathrm{H}}=(1-r_{\mathrm{H}})\mathbf{R}_0+r_{\mathrm{H}}\mathbf{R}_-$. Solving these two linear relations for the endpoint vectors gives
\begin{equation}
\mathbf{R}_0
=
\frac{
r_{\mathrm{H}}\mathbf{R}_{\mathrm{L}}
-
r_{\mathrm{L}}\mathbf{R}_{\mathrm{H}}
}{
r_{\mathrm{H}}-r_{\mathrm{L}}
},
\qquad
\mathbf{R}_-
=
\frac{
(1-r_{\mathrm{L}})\mathbf{R}_{\mathrm{H}}
-
(1-r_{\mathrm{H}})\mathbf{R}_{\mathrm{L}}
}{
r_{\mathrm{H}}-r_{\mathrm{L}}
}.
\label{eq:si_endpoints}
\end{equation}
Thus, the endpoint vectors corresponding to the pure NV$^0$ and NV$^-$ basis spectra can be inferred from two measured mixed spectra and their ZPL-derived contribution ratios, without directly measuring the pure-state spectra.

Once the endpoint vectors have been determined, the NV$^-$ PL contribution ratio of an unknown spectrum is obtained by projecting its mapped vector $\mathbf{R}_{\mathrm{u}}$ onto the spectral manifold defined by $\mathbf{R}_0$ and $\mathbf{R}_-$. Consistent with the linear relation in Eq.~\eqref{eq:si_one_dimensional_manifold}, the estimated contribution ratio is
\begin{equation}
r_{\mathrm{u}}
=
\frac{
(\mathbf{R}_{\mathrm{u}}-\mathbf{R}_0)
\cdot
(\mathbf{R}_--\mathbf{R}_0)
}{
\left\|
\mathbf{R}_--\mathbf{R}_0
\right\|^2
}.
\label{eq:si_projection}
\end{equation}
The combination of the linear spectral mapping, ZPL-based calibration, endpoint determination, and projection therefore converts each measured spectrum into a position along a physically calibrated one-dimensional spectral manifold, from which its NV$^-$ PL contribution ratio is estimated.

\section{Spectral processing and ZPL fitting}
\label{sec:zpl-fitting}

The measured PL spectra were processed by sequentially applying background subtraction, spectral stitching, selection of the analyzed wavelength range, and area normalization. For each spectral segment, a separately acquired background spectrum was subtracted from the measured spectrum. To cover a broad wavelength range, each PL spectrum was acquired in five separate spectral segments, which were stitched after background subtraction to construct a single continuous spectrum. For each pair of adjacent segments, a scale factor was determined by least-squares minimization over the overlapping wavelength region so that their spectral intensities matched as closely as possible. The spectral regions around the NV$^{0}$ and NV$^{-}$ ZPLs were excluded from the scale-factor estimation to prevent the localized ZPL features from affecting the stitching coefficient. The resulting spectra were restricted to the wavelength range of 550--850~nm. This range contains most of the measured PL emission, including the ZPLs and broad phonon sidebands of both NV$^{0}$ and NV$^{-}$. Finally, each spectrum was normalized by its integrated intensity over 550--850~nm and used for the subsequent analysis.

The NV$^{0}$ and NV$^{-}$ ZPL areas required for calibration of the CIE--ZPL method were extracted from the area-normalized spectra in two steps. For each spectrum, a local linear baseline $B_q(\lambda)=b_{0,q}+b_{1,q}\lambda$ was first determined by least-squares fitting to the two side windows of each ZPL. The side windows were 566--570 and 581--585~nm for NV$^{0}$, and 628--632 and 643--647~nm for NV$^{-}$. After subtracting this baseline, the residual intensity was fitted with a Gaussian over 570--581~nm for NV$^{0}$ and 632--643~nm for NV$^{-}$:
\begin{equation}
I(\lambda)-B_q(\lambda)
\simeq
H_q\exp\left[
-\frac{(\lambda-\lambda_{c,q})^2}{2\sigma_q^2}
\right],
\qquad q\in\{0,-\}.
\end{equation}
The peak centers were fixed at the nominal NV$^{0}$ and NV$^{-}$ ZPL wavelengths ($\lambda_{c,0}=575$~nm and $\lambda_{c,-}=637$~nm). The Gaussian amplitudes $H_q$ and widths $\sigma_q$ were obtained by least-squares fitting. The ZPL areas $A_0$ and $A_-$ were taken as the integrated areas of the fitted Gaussian components. For the two calibration spectra, these correspond to $A_{0,\mathrm{H}}$, $A_{-,\mathrm{H}}$, $A_{0,\mathrm{L}}$, and $A_{-,\mathrm{L}}$.

The two spectra used for the CIE--ZPL calibration were selected to enable a stable determination of the sample- and measurement-system-specific coefficient $\alpha$. As shown in Sec.~\ref{sec:geometric_framework_derivation}, $\alpha$ is determined from the ZPL-area differences between the two calibration spectra according to Eq.~\eqref{eq:si_alpha}. Thus, both the numerator and denominator are determined from differences in the respective ZPL areas between the two calibration spectra. If these differences are small, they can become comparable to uncertainties associated with the measurements and ZPL fitting, making the resulting value of $\alpha$ more sensitive to such uncertainties. It is therefore preferable to select two calibration spectra for which the NV$^0$ and NV$^-$ ZPL areas differ sufficiently. In the present measurements, the NV$^-$ PL contribution was expected to vary systematically with excitation power. To obtain calibration spectra with sufficiently different charge-state compositions, the spectra measured at the two ends of the investigated excitation-power range, 1 and 100~mW, were therefore selected. The 1~mW spectrum was used as $S_{\mathrm{H}}(\lambda)$, with the higher NV$^-$ PL contribution, and the 100~mW spectrum as $S_{\mathrm{L}}(\lambda)$, with the lower NV$^-$ PL contribution.

\begin{figure}[htbp]
  \centering
  \includegraphics[width=3.37in]{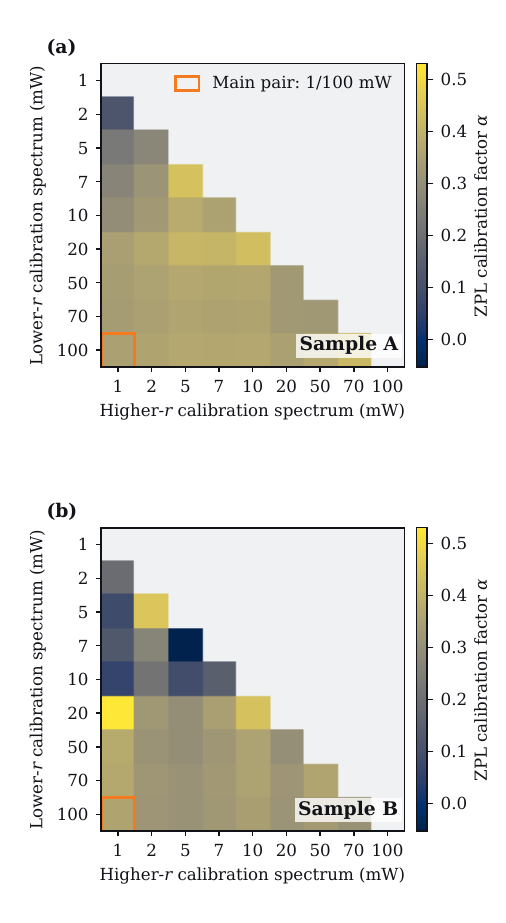}
  \caption{
  Dependence of the ZPL calibration factor $\alpha$ on the choice of calibration spectra.
  Heat maps of $\alpha$ obtained using different combinations of spectra measured at different excitation powers for (a) Sample A and (b) Sample B.
  Each cell represents the value of $\alpha$ calculated from the two spectra measured at the excitation powers indicated on the horizontal and vertical axes.
  The horizontal and vertical axes correspond to $S_{\mathrm{H}}(\lambda)$ and $S_{\mathrm{L}}(\lambda)$, with higher and lower NV$^-$ PL contributions, respectively.
  The same color scale is used for both samples.
  The orange boxes indicate the 1 and 100~mW spectra used as the two calibration spectra in the main analysis.
  }
  \label{fig:calibration_pair_dependence}
\end{figure}

To examine how the choice of calibration spectra affects the determined value of $\alpha$, $\alpha$ was recalculated for combinations of spectra measured at different excitation powers. Figure~\ref{fig:calibration_pair_dependence} shows the resulting $\alpha$ values for Samples A and B. For both samples, some combinations involving spectra measured at nearby excitation powers yielded $\alpha$ values that differed substantially from those obtained for most other combinations, whereas combinations of spectra measured at more widely separated excitation powers tended to yield more similar values of $\alpha$. This behavior is consistent with the discussion above because small differences in the ZPL areas between two calibration spectra make the determined value of $\alpha$ more susceptible to measurement and ZPL-fitting uncertainties. The 1 and 100~mW spectra used in the main analysis yielded $\alpha$ values consistent with those obtained from other combinations spanning a broad excitation-power range.

\section{Comparison methods for charge-state estimation}
\label{sec:reference-methods}

To validate the NV$^-$ PL contribution ratios estimated by the CIE--ZPL method, two comparison methods were used: the dual-excitation protocol (DEP) and ZPL analysis based on Debye--Waller factors (DWF--ZPL). The detailed procedures for each method are described below.

\begin{figure*}[htbp]
    \centering
    \includegraphics[width=6.69in]{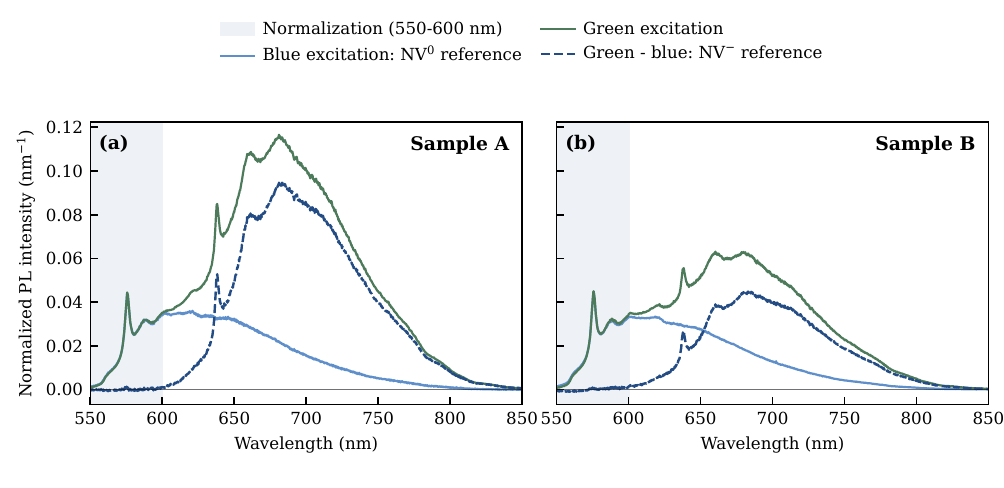}
    \caption{
    Construction of the NV$^0$ and NV$^-$ reference spectra used for the dual-excitation protocol (DEP) for (a) Sample A and (b) Sample B.
    The PL spectrum acquired under 405~nm blue excitation was used as the NV$^0$ reference spectrum, while the 10~mW spectrum acquired under 532~nm green excitation contains contributions from both NV$^0$ and NV$^-$.
    The blue- and green-excitation spectra were independently normalized to the same NV$^0$ contribution over 550--600~nm, indicated by the shaded region.
    The NV$^-$ reference spectrum was obtained by subtracting the normalized blue-excitation spectrum from the normalized green-excitation spectrum.
    }
    \label{fig:dep-reference}
\end{figure*}

For DEP, the PL spectrum acquired under 405~nm blue excitation was used as the NV$^0$ reference spectrum, and an NV$^-$ reference spectrum was constructed in combination with the 10~mW spectrum acquired under 532~nm green excitation~\cite{Thalassinos2025}. The blue- and green-excitation spectra were independently normalized to the same NV$^0$ contribution over 550--600~nm. The normalized blue-excitation spectrum was then subtracted from the normalized green-excitation spectrum to obtain the NV$^-$ reference spectrum. This procedure was performed independently for Samples A and B, as shown in Fig.~\ref{fig:dep-reference}.

Using the resulting NV$^0$ and NV$^-$ reference spectra, each measured spectrum was fitted as a non-negative linear combination of the two reference spectra. Specifically, the measured spectrum $S_i(\lambda)$ was represented over 550--850~nm as
\begin{equation}
S_i(\lambda)
\simeq
c_{0,i}S_{0,\mathrm{ref}}(\lambda)
+
c_{-,i}S_{-,\mathrm{ref}}(\lambda),
\end{equation}
where $c_{0,i}$ and $c_{-,i}$ were determined by non-negative least-squares fitting. The PL contributions $P_{0,i}$ and $P_{-,i}$ were obtained from the fitted coefficients and the integrated intensities of the corresponding reference spectra. The NV$^-$ PL contribution ratio estimated by DEP was then calculated as
\begin{equation}
r_{\mathrm{DEP},i}
=
\frac{P_{-,i}}
{P_{0,i}+P_{-,i}}.
\end{equation}

For the DWF--ZPL method, the total PL contributions from NV$^0$ and NV$^-$ were estimated from their respective ZPL areas using Debye--Waller factors (DWFs), defined as the ratios of ZPL emission to total PL emission. The ZPL areas $A_0$ and $A_-$ were obtained using the fitting procedure described in Sec.~\ref{sec:zpl-fitting}. The Debye--Waller factors
$\mathrm{DWF}_0=e^{-3.3}$ and $\mathrm{DWF}_-=e^{-4.3}$
were used for NV$^0$ and NV$^-$, respectively~\cite{Alsid2019}.
From the relation between the ZPL and total PL intensities, the corresponding total PL contributions were calculated as
\begin{equation}
P_0^{\mathrm{DWF}}
=
\frac{A_0}{\mathrm{DWF}_0}
=
A_0 e^{3.3},
\qquad
P_-^{\mathrm{DWF}}
=
\frac{A_-}{\mathrm{DWF}_-}
=
A_- e^{4.3}.
\label{eq:dwf_total_pl}
\end{equation}

The NV$^-$ PL contribution ratio obtained by the DWF--ZPL method was therefore calculated from the corresponding total PL contributions as
\begin{equation}
r_{\mathrm{DWF-ZPL}}
=
\frac{
A_- e^{4.3}
}{
A_0 e^{3.3} + A_- e^{4.3}
}.
\label{eq:dwf_ratio}
\end{equation}

This procedure was applied to each measured PL spectrum and the resulting values were used for comparison with the CIE--ZPL and DEP estimates.

\section{Validation of the one-dimensional manifold assumption}
\label{sec:manifold-validation}

To test whether the approximately one-dimensional behavior observed after mapping into CIE XYZ space is already present in the measured spectral variation, principal component analysis (PCA) was performed on the area-normalized PL spectra. In addition, the orthogonal deviation of each mapped CIE XYZ vector from the calibrated spectral manifold was evaluated to quantify the extent to which the measured vectors follow the one-dimensional manifold.

\begin{figure*}[htbp]
    \centering
    \includegraphics[width=6.69in]{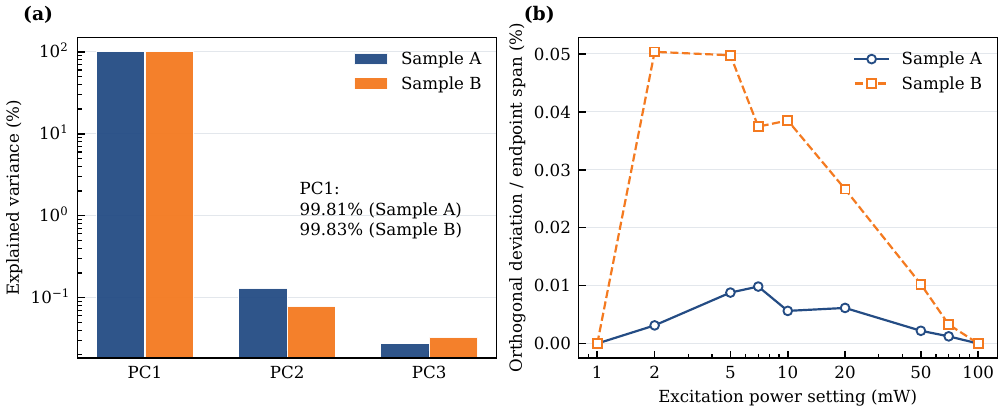}
    \caption{
    Validation of the one-dimensional spectral manifold assumption.
    (a) Explained variance obtained from principal component analysis of the area-normalized PL spectra for Samples A and B. The first principal component accounts for 99.81\% and 99.83\% of the spectral variance for Samples A and B, respectively.
    (b) Orthogonal deviation of each mapped CIE XYZ vector from the calibrated spectral manifold, normalized by the span between the endpoints $\mathbf{R}_0$ and $\mathbf{R}_-$. The deviations remain within approximately 0.05\% of the endpoint span over the investigated excitation-power settings. The deviations at 1 and 100~mW are approximately zero because these spectra were used as the calibration pair defining the spectral manifold.
    }
    \label{fig:manifold_validation}
\end{figure*}

The PCA showed that the first principal component accounted for 99.81\% and 99.83\% of the spectral variance for Samples A and B, respectively [Fig.~\ref{fig:manifold_validation}(a)]. The orthogonal deviations from the calibrated spectral manifold remained within approximately 0.05\% of the endpoint span over the investigated excitation-power settings [Fig.~\ref{fig:manifold_validation}(b)]. These results indicate that the excitation-power-dependent spectral variation is effectively described by a single degree of freedom and support the one-dimensional spectral manifold assumed in the present framework.

\section{Noise robustness and additional validation}
\label{sec:noise-validation}

This section provides additional details of the noise-perturbation analysis used to evaluate the robustness of charge-state estimation to spectral noise, together with supplementary validation of the resulting estimates. In this analysis, noise was added only to the spectrum to be estimated, while all method-specific calibration inputs were held fixed. For CIE--ZPL, the endpoint vectors $\mathbf{R}_{0}$ and $\mathbf{R}_{-}$ determined from the calibration spectra were fixed throughout the simulation. Likewise, the NV$^{0}$ and NV$^{-}$ reference spectra constructed for DEP were held fixed, while the literature-derived parameters used for DWF--ZPL were unchanged. The PL spectrum acquired from Sample A at a nominal laser-output setting of 5~mW was used as the test spectrum.

The noise level was defined based on the root-mean-square intensity of the noise-free spectrum. Zero-mean white Gaussian noise was added with a noise standard deviation given by
\begin{equation}
\sigma_{n}
=
\frac{I_{\mathrm{RMS}}}{\mathrm{SNR}},
\end{equation}
where SNR is the signal-to-noise ratio and $I_{\mathrm{RMS}}$ is the root-mean-square intensity of the noise-free spectrum, given by
\begin{equation}
I_{\mathrm{RMS}}
=
\left[
\frac{1}{N}
\sum_{j=1}^{N} I_{j}^{2}
\right]^{1/2},
\end{equation}
$I_j$ is the PL intensity of the noise-free spectrum at the $j$th wavelength point, and $N$ is the number of wavelength points. After noise addition, each noise-perturbed spectrum was area-normalized again over 550--850~nm for CIE--ZPL estimation. DEP and DWF--ZPL were applied directly to the noise-perturbed spectrum without this renormalization. The SNR was varied from 5 to 50 in increments of 5, and 1000 Monte Carlo trials were performed at each SNR. Within each trial, the same noise realization was used for the CIE--ZPL, DEP, and DWF--ZPL methods, and the NV$^{-}$ PL contribution ratio was estimated using each method.

\begin{figure*}[htbp]
    \centering
    \includegraphics[width=6.69in]{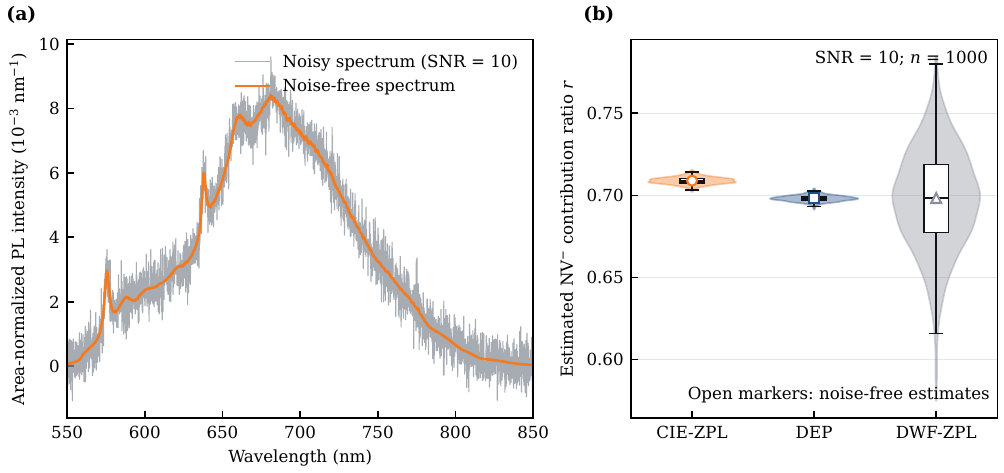}
    \caption{
    Representative noise-perturbed PL spectrum and distributions of NV$^{-}$ PL contribution-ratio estimates under additive spectral noise.
    (a) Noise-free PL spectrum of Sample A acquired at a nominal laser-output setting of 5~mW and a representative spectrum obtained after adding white Gaussian noise at $\mathrm{SNR}=10$.
    (b) Distributions of the NV$^{-}$ PL contribution-ratio estimates obtained using CIE--ZPL, DEP, and DWF--ZPL from 1000 Monte Carlo trials at $\mathrm{SNR}=10$.
    Open markers indicate the estimates obtained from the noise-free spectrum.
    }
    \label{fig:noise_distributions}
\end{figure*}

To illustrate the effect of the noise perturbation, Fig.~\ref{fig:noise_distributions} shows a representative noise-perturbed spectrum and the corresponding distributions of the estimated NV$^{-}$ PL contribution ratios. Figure~\ref{fig:noise_distributions}(a) compares the noise-free spectrum with a representative spectrum obtained after adding white Gaussian noise at $\mathrm{SNR}=10$. Figure~\ref{fig:noise_distributions}(b) shows the distributions of the estimated NV$^{-}$ PL contribution ratios obtained from 1000 Monte Carlo trials at the same SNR for the CIE--ZPL, DEP, and DWF--ZPL methods, with open markers indicating the estimates obtained from the noise-free spectrum. The CIE--ZPL and DEP estimates are distributed over relatively narrow ranges, whereas DWF--ZPL exhibits a substantially broader distribution.

\begin{figure}[htbp]
    \centering
    \includegraphics[width=3.37in]{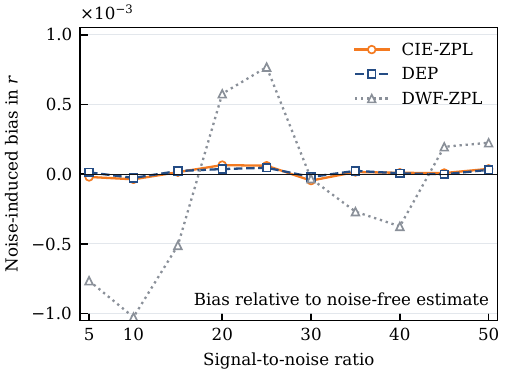}
    \caption{
    SNR dependence of the noise-induced bias in NV$^{-}$ PL contribution-ratio estimation under additive spectral noise.
    For CIE--ZPL, DEP, and DWF--ZPL, the bias at each SNR is defined as the difference between the mean estimate over the Monte Carlo trials and the estimate obtained from the noise-free spectrum.
    }
    \label{fig:noise_bias}
\end{figure}

To further examine whether random spectral noise introduces a systematic shift in the estimated contribution ratio, the SNR dependence of the noise-induced bias was evaluated for each method. The bias at each SNR was defined as
\begin{equation}
\mathrm{bias}
=
\left\langle
r_{\mathrm{noisy}}
\right\rangle
-
r_{\mathrm{nf}},
\end{equation}
where $\left\langle r_{\mathrm{noisy}} \right\rangle$ is the mean estimate over the Monte Carlo trials and $r_{\mathrm{nf}}$ is the estimate obtained by applying the same method to the noise-free spectrum. As shown in Fig.~\ref{fig:noise_bias}, the biases of CIE--ZPL and DEP remain close to zero over the investigated SNR range. Although DWF--ZPL shows larger fluctuations, particularly at low SNR, the absolute bias remains small overall, indicating that random spectral noise does not introduce a pronounced systematic shift in the estimated contribution ratio.

\section{Initialization dependence of non-negative matrix factorization}
\label{sec:nnmf-initialization}

To evaluate the dependence of the NNMF solution on initialization, two-component NNMF was applied to the nine PL spectra measured from Sample A at different excitation powers. NNMF was performed using an alternating least-squares algorithm, and two initialization conditions were compared: random initialization and 1/100~mW initialization, in which the measured 1 and 100~mW spectra were used as the initial basis spectra.

\begin{figure*}[htbp]
    \centering
    \includegraphics[width=6.69in]{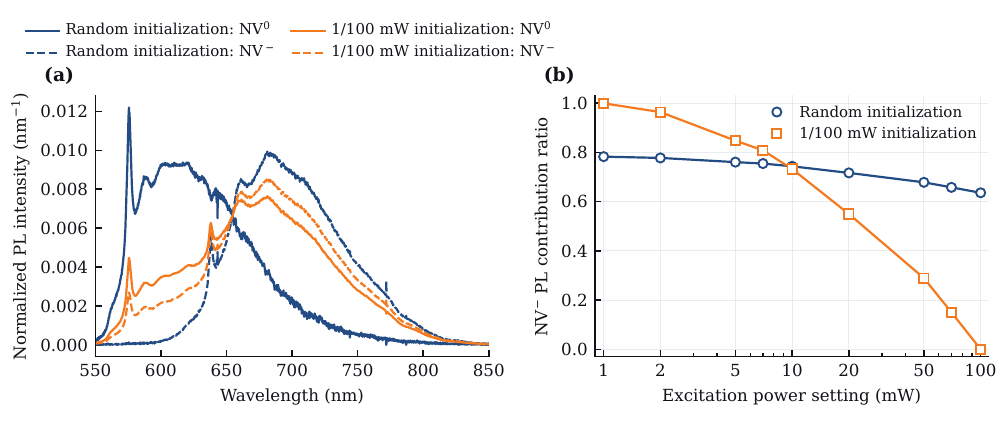}
    \caption{
    Dependence of the NNMF results on initialization.
    (a) Basis spectra obtained using random initialization and 1/100~mW initialization based on the measured 1 and 100~mW spectra. Solid and dashed lines indicate the NV$^{0}$ and NV$^{-}$ components, respectively.
    (b) Excitation-power dependence of the NV$^{-}$ PL contribution ratio obtained from each initialization.
    }
    \label{fig:nnmf_initialization_dependence}
\end{figure*}

The NNMF results obtained with the two initialization conditions are shown in Fig.~\ref{fig:nnmf_initialization_dependence}. Random initialization and 1/100~mW initialization yielded clearly different shapes of the extracted basis spectra [Fig.~\ref{fig:nnmf_initialization_dependence}(a)]. This difference was also reflected in the estimated contributions of the two components, resulting in substantially different NV$^{-}$ PL contribution ratios between the two initialization conditions [Fig.~\ref{fig:nnmf_initialization_dependence}(b)].

Despite yielding different basis spectra and NV$^{-}$ PL contribution ratios, the two NNMF solutions reconstructed the measured spectra with comparable accuracy. The root-mean-square (RMS) reconstruction errors were $1.32\times10^{-5}~\mathrm{nm}^{-1}$ for random initialization and $1.11\times10^{-5}~\mathrm{nm}^{-1}$ for 1/100~mW initialization, both of the same order of magnitude. Thus, the reconstruction error alone did not provide a basis for selecting between the two solutions.

These results show that accurate spectral reconstruction alone does not guarantee the physical identifiability of the spectral components extracted by NNMF or their contributions. For the same experimental dataset, different initializations can yield different basis spectra and NV$^{-}$ PL contribution ratios while producing comparable reconstruction errors. Therefore, interpreting the components obtained by NNMF as physical NV$^{0}$ and NV$^{-}$ contributions requires physical information or calibration independent of the spectral reconstruction itself.

\section{Fisher information and Cram\'er--Rao bounds}
\label{sec:fisher-information}

To quantify the information about $r$ retained by each spectral representation, we calculated the Fisher information (FI) for the full spectrum, CIE XYZ coordinates, and two ZPL areas. The analysis used Sample A and the noise model of Sec.~\ref{sec:noise-validation}. All basis spectra and calibration parameters were held fixed, with $r$ as the only unknown parameter. We first establish the full-spectrum FI as a reference, then derive the FI of the CIE XYZ and ZPL representations, and finally compare their information retention and corresponding Cram\'er--Rao bounds.

\subsection{Fisher information of the full spectrum}

To establish a common basis for comparing the three representations, we first formulate the full-spectrum observation model using the two-component mixture of Sec.~\ref{sec:geometric_framework_derivation} and the additive-noise model of Sec.~\ref{sec:noise-validation}. Let $\mathbf{S}_0$ and $\mathbf{S}_-$ denote the discrete counterparts of $S_0(\lambda)$ and $S_-(\lambda)$, sampled at $N$ wavelengths over 550--850~nm. For this FI comparison, the basis spectra were constructed using DEP as described in Sec.~\ref{sec:reference-methods} and individually normalized to unit area over the analysis range. With $\boldsymbol\mu(r)$ denoting the mean spectrum and $\boldsymbol\epsilon$ the additive noise, the observed spectral-intensity vector $\mathbf{x}$ is modeled as
\begin{equation}
\begin{aligned}
\mathbf{x}&=\boldsymbol\mu(r)+\boldsymbol\epsilon,\\
\boldsymbol\mu(r)&=(1-r)\mathbf{S}_0+r\mathbf{S}_-,\\
\boldsymbol\epsilon&\sim\mathcal{N}(\mathbf{0},\sigma_n^2\mathbf{I}_N),
\end{aligned}
\label{eq:si_fi_observation}
\end{equation}
where $\mathbf{I}_N$ is the $N\times N$ identity matrix and $\sigma_n=I_{\mathrm{RMS}}/\mathrm{SNR}$ is evaluated from the 5~mW test spectrum and held fixed when differentiating with respect to $r$.

The FI can be evaluated from the sensitivity of the mean to $r$ and the observation covariance. For a Gaussian observation $\mathbf{y}$ with mean $\boldsymbol\mu_y(r)$ and covariance $\boldsymbol\Sigma_y$ independent of $r$, it is given by
\begin{equation}
\mathcal{I}_y
=\mathbb{E}\!\left[\left(\frac{\partial\ln p(\mathbf{y}\mid r)}{\partial r}\right)^2\right]
=\mathbf{d}_y^{\mathrm{T}}\boldsymbol\Sigma_y^{-1}\mathbf{d}_y,
\label{eq:si_fi_gaussian}
\end{equation}
where $p(\mathbf{y}\mid r)$ is the conditional probability density of the observation, $\mathbb{E}$ denotes expectation with respect to this density, and $\mathbf{d}_y=\partial\boldsymbol\mu_y/\partial r$ is the mean sensitivity.

For the full spectrum, the sensitivity is $\mathbf{d}=\mathbf{S}_--\mathbf{S}_0$, giving
\begin{equation}
\mathcal{I}_{\mathrm{Full}}
=\frac{\mathbf{d}^{\mathrm{T}}\mathbf{d}}{\sigma_n^2}.
\label{eq:si_fi_full}
\end{equation}

\subsection{Fisher information of the CIE XYZ representation}

To evaluate the FI retained by the CIE XYZ representation, we first propagate the spectral noise through area normalization and CIE mapping to obtain the XYZ covariance, and then determine the mean sensitivity to $r$.

To describe this normalization, let $\mathbf{a}$ contain the trapezoidal integration weights, so that the spectral area is
$\mathcal{A}(\mathbf{x})=\mathbf{a}^{\mathrm{T}}\mathbf{x}$ and the normalized spectrum is
$\mathbf{x}_{\mathrm{norm}}(\mathbf{x})=\mathbf{x}/\mathcal{A}(\mathbf{x})$.
Using $j$ and $\ell$ to index wavelength points, differentiating each component
$x_{\mathrm{norm},j}=x_j/\mathcal{A}(\mathbf{x})$ with respect to $x_\ell$ gives
$
\frac{\partial x_{\mathrm{norm},j}}{\partial x_\ell}
=
\frac{\delta_{j\ell}}{\mathcal{A}}
-
\frac{x_j a_\ell}{\mathcal{A}^2},
$
where $\delta_{j\ell}$ is the Kronecker delta.
Let $\mathbf{x}_*$ denote the measured 5~mW test spectrum before noise addition, and define its spectral area as
$\mathcal{A}_*=\mathcal{A}(\mathbf{x}_*)=\mathbf{a}^{\mathrm{T}}\mathbf{x}_*$.
The Jacobian of the normalization with respect to the spectral intensities, evaluated at $\mathbf{x}_*$, is then
\begin{equation}
\mathbf{J}_*
=
\left.
\frac{\partial\mathbf{x}_{\mathrm{norm}}}
{\partial\mathbf{x}}
\right|_{\mathbf{x}_*}
=
\frac{\mathbf{I}_N}{\mathcal{A}_*}
-
\frac{\mathbf{x}_*\mathbf{a}^{\mathrm{T}}}
{\mathcal{A}_*^2}.
\label{eq:si_fi_normalization}
\end{equation}
The first-order Taylor expansion provides a linear approximation to the effect of additive noise on the normalized spectrum:
\begin{equation}
\begin{aligned}
\mathbf{x}_{\mathrm{norm}}(\mathbf{x}_*+\boldsymbol\epsilon)
&\simeq
\mathbf{x}_{\mathrm{norm}}(\mathbf{x}_*)
+
\mathbf{J}_*\boldsymbol\epsilon,\\
\boldsymbol\Sigma_{\mathrm{norm}}
&\simeq
\mathbf{J}_*\boldsymbol\Sigma_x\mathbf{J}_*^{\mathrm{T}}
=
\sigma_n^2\mathbf{J}_*\mathbf{J}_*^{\mathrm{T}},
\end{aligned}
\label{eq:si_fi_normalized_covariance}
\end{equation}
where $\boldsymbol\Sigma_x=\sigma_n^2\mathbf{I}_N$ is the covariance before normalization and
$\boldsymbol\Sigma_{\mathrm{norm}}$ is the covariance after normalization.

The normalized spectrum is then mapped to the CIE XYZ vector $\mathbf{R}=(X,Y,Z)^{\mathrm{T}}=\mathbf{W}_{\mathrm{CIE}}\mathbf{x}_{\mathrm{norm}}$, the discrete form of the linear mapping in Sec.~\ref{sec:geometric_framework_derivation}. The $3\times N$ mapping matrix has entries $(\mathbf{W}_{\mathrm{CIE}})_{kj}=a_j w_k(\lambda_j)$, where $k\in\{1,2,3\}$ indexes the XYZ components, $\lambda_j$ is the $j$th wavelength, and $(w_1,w_2,w_3)=(\overline{x},\overline{y},\overline{z})$ are the CIE 1931 color-matching functions linearly interpolated onto the spectral wavelength grid. Propagating the normalized spectral covariance through this linear mapping gives
\begin{equation}
\begin{aligned}
\boldsymbol\Sigma_{\mathrm{CIE}}
&=\mathbf{W}_{\mathrm{CIE}}\boldsymbol\Sigma_{\mathrm{norm}}\mathbf{W}_{\mathrm{CIE}}^{\mathrm{T}}\\
&\simeq\sigma_n^2\mathbf{W}_{\mathrm{CIE}}\mathbf{J}_*
\mathbf{J}_*^{\mathrm{T}}\mathbf{W}_{\mathrm{CIE}}^{\mathrm{T}}.
\end{aligned}
\label{eq:si_fi_cie_covariance}
\end{equation}

This covariance is evaluated at the measured 5~mW spectrum and held fixed when calculating the FI. The remaining quantity is the mean sensitivity, which is obtained from the DEP basis spectra.

To obtain the mean sensitivity, we use the normalization of the basis spectra, $\mathbf{a}^{\mathrm{T}}\mathbf{S}_0=\mathbf{a}^{\mathrm{T}}\mathbf{S}_-=1$. Consequently, $\mathbf{a}^{\mathrm{T}}\boldsymbol\mu(r)=(1-r)+r=1$, so varying $r$ changes the spectral shape but not its integrated area. Normalization therefore leaves the noise-free mixture unchanged, $\mathbf{x}_{\mathrm{norm}}(\boldsymbol\mu(r))=\boldsymbol\mu(r)$, and differentiating its CIE representation gives
\begin{equation}
\mathbf{d}_{\mathrm{CIE}}
=
\frac{\partial}{\partial r}
\left\{
\mathbf{W}_{\mathrm{CIE}}
\mathbf{x}_{\mathrm{norm}}(\boldsymbol\mu(r))
\right\}
=
\mathbf{W}_{\mathrm{CIE}}\mathbf{d}.
\label{eq:si_fi_cie_sensitivity}
\end{equation}
Combining this sensitivity with the propagated covariance yields
\begin{equation}
\mathcal{I}_{\mathrm{CIE}}
=\mathbf{d}_{\mathrm{CIE}}^{\mathrm{T}}
\boldsymbol\Sigma_{\mathrm{CIE}}^{-1}\mathbf{d}_{\mathrm{CIE}}.
\label{eq:si_fi_cie}
\end{equation}

To validate the first-order analytical covariance used in this FI calculation, we performed Monte Carlo simulations using the same additive white Gaussian noise model and SNR values as in the noise-robustness analysis of Sec.~\ref{sec:noise-validation}. A common set of 10,000 standard normal noise realizations was generated and reused across all SNR values, with each realization scaled by $\sigma_n = I_{\mathrm{RMS}}/\mathrm{SNR}$ before being added to the measured 5~mW test spectrum. Each noise-perturbed spectrum was then area-normalized and mapped to CIE XYZ without linearization. The empirical covariance matrix of the resulting XYZ vectors, denoted $\boldsymbol{\Sigma}_{\mathrm{CIE}}^{\mathrm{MC}}$, was compared with the analytical covariance matrix $\boldsymbol{\Sigma}_{\mathrm{CIE}}$ in Eq.~\eqref{eq:si_fi_cie_covariance} using the relative Frobenius-norm difference, $\|\boldsymbol{\Sigma}_{\mathrm{CIE}}^{\mathrm{MC}}-\boldsymbol{\Sigma}_{\mathrm{CIE}}\|_F/\|\boldsymbol{\Sigma}_{\mathrm{CIE}}\|_F$, where $\|\cdot\|_F$ denotes the Frobenius norm. Across the SNR range of 5--50, the relative Frobenius-norm difference ranged from 0.290\% to 0.294\%. This close agreement supports the accuracy of the first-order covariance propagation for the test spectrum over the investigated noise range.

\subsection{Fisher information of the ZPL observables}

To evaluate the information retained by the two ZPL areas, we express background subtraction and area extraction as a linear transformation of the spectrum, then use this transformation to obtain the area sensitivities and covariance. The following construction applies to each charge state $q\in\{0,-\}$, with the peak parameters denoted as in Sec.~\ref{sec:zpl-fitting}. The peak centers and widths were fixed for this linear transformation. The centers were $\lambda_{c,0}=575$~nm and $\lambda_{c,-}=637$~nm. For the Gaussian widths, the median of the fitted $\sigma_q$ values across the nine excitation-power settings was used as a fixed width for each charge state.
These fixed median widths are denoted by $\sigma_{q,\mathrm{med}}$, with $\sigma_{0,\mathrm{med}}=1.76$~nm and $\sigma_{-,\mathrm{med}}=1.87$~nm. For NV$^0$, the background windows were 566--570 and 581--585~nm, and the peak window was 570--581~nm. For NV$^-$, the corresponding windows were 628--632 and 643--647~nm for the background and 632--643~nm for the peak.

We first construct the linear operation that subtracts the fitted background from each peak window. Let $\mathbf{P}_{q,b}$ and $\mathbf{P}_{q,f}$ select the background and peak-fitting windows, respectively, with $b$ and $f$ labeling the two window types. As in Sec.~\ref{sec:zpl-fitting}, the local background is modeled as $B_q(\lambda)=b_{0,q}+b_{1,q}\lambda$, where $b_{0,q}$ and $b_{1,q}$ are the intercept and slope. Writing $\mathbf{b}_q=(b_{0,q},b_{1,q})^{\mathrm{T}}$, the background sampled over all wavelengths is $\mathbf{M}\mathbf{b}_q$. The $j$th row of the $N\times2$ design matrix is $\mathbf{M}_{j,:}=(1,\lambda_j)$. Its two columns multiply the intercept and slope, respectively.

The matrix $\mathbf{M}_{q,b}=\mathbf{P}_{q,b}\mathbf{M}$ contains the rows in the background windows. A least-squares fit to the measured intensities $\mathbf{P}_{q,b}\mathbf{x}$ gives $\widehat{\mathbf{b}}_q=\mathbf{M}_{q,b}^{+}\mathbf{P}_{q,b}\mathbf{x}$, where $+$ denotes the Moore--Penrose pseudoinverse. Selecting the peak-window rows gives $\mathbf{M}_{q,f}=\mathbf{P}_{q,f}\mathbf{M}$, so $\mathbf{M}_{q,f}\widehat{\mathbf{b}}_q$ evaluates the fitted background within the peak window. Subtracting this background gives the corrected peak signal
\begin{equation}
\begin{aligned}
\mathbf{y}_{q,f}
&=\mathbf{P}_{q,f}\mathbf{x}-\mathbf{M}_{q,f}\widehat{\mathbf{b}}_q\\
&=\left(\mathbf{P}_{q,f}-\mathbf{M}_{q,f}\mathbf{M}_{q,b}^{+}\mathbf{P}_{q,b}\right)\mathbf{x}.
\end{aligned}
\label{eq:si_fi_zpl_background}
\end{equation}
The next step converts the corrected peak signal into an integrated ZPL area. Let $\mathbf{g}_q$ contain the fixed Gaussian profile values $g_{q,j}
=
\exp\left[
-\frac{(\lambda_j-\lambda_{c,q})^2}
{2\sigma_{q,\mathrm{med}}^2}
\right]$ at the wavelengths in the peak window, ordered as selected by $\mathbf{P}_{q,f}$. The least-squares amplitude is $\widehat H_q=\mathbf{g}_q^{\mathrm{T}}\mathbf{y}_{q,f}/(\mathbf{g}_q^{\mathrm{T}}\mathbf{g}_q)$. Multiplication by $\sqrt{2\pi}\,\sigma_{q,\mathrm{med}}$ converts this amplitude to the estimated integrated Gaussian area, which can be written using an area-weight vector $\mathbf{w}_{A,q}$ as
\begin{equation}
\begin{aligned}
\widehat A_q&=\mathbf{w}_{A,q}^{\mathrm{T}}\mathbf{x},\\
\mathbf{w}_{A,q}^{\mathrm{T}}
&=\sqrt{2\pi}\,\sigma_{q,\mathrm{med}}\,
\frac{\mathbf{g}_q^{\mathrm{T}}}{\mathbf{g}_q^{\mathrm{T}}\mathbf{g}_q}
\left(\mathbf{P}_{q,f}-\mathbf{M}_{q,f}\mathbf{M}_{q,b}^{+}\mathbf{P}_{q,b}\right).
\end{aligned}
\label{eq:si_fi_zpl_weights}
\end{equation}
To obtain the joint sensitivity and covariance of the two areas, we stack $\mathbf{w}_{A,0}^{\mathrm{T}}$ and $\mathbf{w}_{A,-}^{\mathrm{T}}$ as the first and second rows of the $2\times N$ matrix $\mathbf{W}_{\mathrm{ZPL}}$. Thus, $\mathbf{W}_{\mathrm{ZPL}}\mathbf{x}=(\widehat A_0,\widehat A_-)^{\mathrm{T}}$. Applying this transformation to the full-spectrum sensitivity and noise covariance gives the ZPL FI as
\begin{equation}
\begin{aligned}
\mathbf{d}_{\mathrm{ZPL}}&=\mathbf{W}_{\mathrm{ZPL}}\mathbf{d},\\
\boldsymbol\Sigma_{\mathrm{ZPL}}
&=\sigma_n^2\mathbf{W}_{\mathrm{ZPL}}\mathbf{W}_{\mathrm{ZPL}}^{\mathrm{T}},\\
\mathcal{I}_{\mathrm{ZPL}}
&=\mathbf{d}_{\mathrm{ZPL}}^{\mathrm{T}}
\boldsymbol\Sigma_{\mathrm{ZPL}}^{-1}\mathbf{d}_{\mathrm{ZPL}}.
\end{aligned}
\label{eq:si_fi_zpl}
\end{equation}
\subsection{Information retention and Cram\'er--Rao lower bounds}

Having obtained the FI for the three observation models, we compare the information retained by the CIE XYZ and ZPL representations relative to the full spectrum. The resulting FI ratios were $\mathcal{I}_{\mathrm{CIE}}/\mathcal{I}_{\mathrm{Full}}=0.81$ and $\mathcal{I}_{\mathrm{ZPL}}/\mathcal{I}_{\mathrm{Full}}=0.0080$, corresponding to information retention of 81\% and 0.80\%, respectively.

To relate these information measures to estimation precision, we use the Cram\'er--Rao lower bound (CRLB), which provides a lower bound on the variance of an unbiased estimator under the usual regularity conditions. For each observation model, the variance bound and the corresponding standard-deviation bound plotted in the main text were calculated as
\begin{equation}
\operatorname{Var}(r)\geq\frac{1}{\mathcal{I}_y},
\qquad
\sigma_{r,\mathrm{CRLB}}=\frac{1}{\sqrt{\mathcal{I}_y}}.
\label{eq:si_fi_crlb}
\end{equation}
Because all covariances scale as $\sigma_n^2$ and the sensitivities are fixed, the FI scales as $\mathrm{SNR}^2$, the standard-deviation CRLB scales as $\mathrm{SNR}^{-1}$, and the retention ratios are independent of SNR within this model.

\end{document}